\documentstyle[aps,epsf,twocolumn]{revtex}
\def\la{\langle} 
\def\ra{\rangle} 
\def\be{\begin{eqnarray}} 
\def\ee{\end{eqnarray}}

\newcommand{\eq}{\begin{equation}} \newcommand{\eqx}{\end{equation}}

\newcommand{\eqn}{\begin{eqnarray}} \newcommand{\eqnx}{\end{eqnarray}}
\newcommand{\f}[2]{\frac{#1}{#2}}
\newcommand{\lra}{\longrightarrow}
\newcommand{\Tr}{\mbox{\rm Tr}}

\newcommand{\lm}{\lambda}

\newcommand{\arr}[4]{
\left(\begin{array}{cc}
#1&#2\\
#3&#4
\end{array}\right)
}

\newcommand{\Qt}{\tilde{Q}}

\begin{document}
\draft

\title{\bf Critical Scaling at Zero Virtuality in QCD}

\author{ {\bf Romuald A. Janik}$^1$, {\bf Maciej A.  Nowak}$^1$ ,
{\bf G\'{a}bor Papp}$^{2}$ and {\bf Ismail Zahed}$^3$}

\address{$^1$ Department of Physics, Jagellonian University, 30-059
Krakow, Poland.
\\ $^2$ITP, Univ. Heidelberg, Philosophenweg 19, D-69120 Heidelberg, 
	Germany \& \\ Institute for
Theoretical Physics, E\"{o}tv\"{o}s University, Budapest, Hungary\\
$^3$Department of Physics and Astronomy, SUNY, Stony Brook, New York
11794, USA.} 
\date{\today} \maketitle

\begin{abstract}
We show that at the critical point of chiral random matrix models,
novel scaling laws for the inverse moments of the eigenvalues are 
expected. We evaluate explicitly the pertinent microscopic spectral 
density, and find it in agreement with numerical calculations.
We suggest that similar sum rules are of relevance to QCD at the
critical temperature, and even above if the transition is amenable
to a Ginzburg-Landau description.

\end{abstract}

{\bf 1.}
A large number of physical phenomena can be modeled using random matrix
models~\cite{GENERAL}. An important aspect of these models is their ability
to capture the generic form of spectral correlations in the ergodic regime 
of quantum systems. This regime is reached by electrons traveling a long 
time in disordered metallic grains~\cite{METAL} or virtual quarks moving a 
long proper time in a small Euclidean volume~\cite{QCD}.

In QCD, the ergodic regime is 
characterized by a huge accumulation of quarks eigenvalues near zero 
virtuality. This is best captured by the Banks-Casher~\cite{BANKS} relation 
$|\la\overline{q} q\ra |\equiv\Sigma=\pi\rho (0)$, where the nonvanishing of the 
chiral condensate in the vacuum signals a finite quark density 
$\rho (\lambda =0)\neq 0$ at zero virtuality. This behavior is
at the origin of spectral sum rules~\cite{SMILGA}, which are reproduced
by chiral random matrix models~\cite{SHURYAK}. These sum rules reflect 
on the distribution of quark eigenvalues and correlations~\cite{VERZA}.

If QCD is to undergo a second or higher order chiral transition, then at
the critical point there is a dramatic reorganization of the light quark 
states near zero virtuality as the quark condensate vanishes. In section
2, we suggest that such a reorganization is followed by new scaling laws,
which are captured by a novel microscopic limit. In section 3, we use a
chiral random matrix model with a mean-field transition to illustrate our
point. In section 4, we explicitly construct the pertinent microscopic
spectral distribution in the quenched case and compare it to numerical
calculations. In sections 5,6 we argue that the sum rules following
from the new scaling, may be of relevance
to QCD at the chiral critical point, provided that the transition follows
the general lore of universality, and suggest that  spectral
correlations persist even above the transition temperature.

{\bf 2.}
The nonvanishing of $\Sigma$ in the QCD vacuum implies that the number
of quark states in a volume $V$, $N(E)=V\int\!d\lambda \rho(\lambda )$ in the
virtuality band $E$ around $0$ grows linearly 
with $E$, that is $N(E)\sim EV$. As a result, the level spacing 
$\Delta=dE/dN\sim 1/V$ for $N\sim 1$, and the eigenvalues of the quark
operator obey spectral sum rules~\cite{SMILGA}. 
During a second or higher order phase
transition $\Sigma$ vanishes in the chiral limit. Scaling arguments give
$\Sigma\sim m^{1/\delta}$ at the critical point, where $m$ is the 
current quark mass \cite{KOCIC}. It follows
again from the Banks-Casher relation~\cite{BANKS}, that for small virtualities
$\lambda$, $\rho (\lambda )\sim |\lambda|^{1/\delta}$ to leading 
order in the 
current quark mass~\cite{USNJL}. 
Hence, $N(E)\sim VE^{1+1/\delta}$, and the level 
spacing is now $\Delta_*\sim 1/V^{\delta/(\delta +1)}$ at $N\sim 1$.

For a mean-field exponent $\delta=3$, and we have $\Delta_*=V^{-3/4}$, which 
is intermediate between 
$V^{-1}$ in the spontaneously broken phase and $V^{-1/4}$ 
in free space. At the critical point there are still level correlations in
the quark spectrum except for the free limit,  corresponding formally
to $\delta=1/3$.
We now conjecture that at the critical point, the rescaling of the quark
eigenvalues through $\lambda\rightarrow\lambda/\Delta_*$, yields new spectral
sum rules much like the rescaling with $\Delta=1/V$ in the 
vacuum~\cite{SMILGA}. The master
formula for the diagonal sum rules is given by the (dimensionless)
microscopic density of states
\be
\nu_* (s) = \lim_{V\to\infty}\, (V\Delta_*) \, \rho (s\Delta_*)
\label{1}
\ee
and similarly for the off-diagonal sum rules in terms of the microscopic
multi-level correlators. Since we lack an accurate effective action 
formulation of QCD at $T=T_c$ (a possibility based on universality is
discussed below), the nature and character of these sum rules is not 
{\it a priori} known, but could easily be established using lattice 
simulations in QCD. 

Could these sum rules be shared by random matrix models? 
We will postpone the answer to this question till section 5, and 
instead show in what follows that the present scaling laws hold at zero 
virtuality for chiral random matrix models with mean-field exponents.

\vskip 0.3cm

{\bf 3.} 
Consider the set of chiral random plus deterministic matrices

\be
{\cal M} = \arr{im}{t+A}{t+A^\dagger}{im}
\label{2}
\ee
where $A$ is an $N\times N$ complex matrix with Gaussian weight,
$m$ a `mass' parameter, and $t$ a `temperature' parameter. Such
matrices or variant thereof have been investigated by a number of 
authors in the recent past~\cite{ALL}. Their associated density of 
states is

\be
\rho (\lambda ) =\frac 1{2N} \la {\rm Tr} \,\delta (\lambda -\cal M )\ra
\label{3}
\ee
where the averaging is carried using the weight associated to the
following partition function
\eq
Z [m,t]=\int\!dA dA^\dagger\, \det{\cal M}^{N_f} e^{-N\Tr AA^\dagger}  \,.
\label{4}
\eqx
For $t=m=0$ the density of states is $\rho (0)=1/\pi$, while zero for
$t\geq 1$ and $m=0$. At $t=1$, $\rho(\lambda 
)=|\lambda |^{1/3}$, which indicates that the $t$-driven transition 
is mean-field with $\rho(0)$ as an order parameter. In particular, at
$t=1$ the level spacing is $\Delta_*=N^{-3/4}$ near zero.

Standard bosonization of the partition function (\ref{4}) yields
\eq
Z [m,t]=\int\!\!dP dP^\dagger\,
e^{N\log\det\!\big[(\!m\!+\!P)(m\!+\!P^\dagger)+t^2\big]\!
	-\!N\Tr PP^\dagger}
\label{5}
\eqx
where $P$ is an $N_f\times N_f$ complex matrix.
We may shift $P$ by the mass matrix $P\lra Q=P+m$ and get
\be
\label{e.last}
Z [m,t] =\int\!dQ dQ^\dagger\, \det(QQ^\dagger+t^2)^N \\
	\times e^{-N\big[\Tr QQ^\dagger -m\Tr(Q+Q^\dagger) +m^2\big]} 
	\nonumber
\label{6}
\ee
dropping the irrelevant normalization factor.
We will now specialize to the critical temperature $t=1$, and denote the
rescaled mass by $x=im/\Delta_*$ and eigenvalue by $s=\lambda/\Delta_*$.
This suggests the rescaling $Q\lra \Qt=N^{1/4} Q$, so that
\eqn
Z [x] = \int\!d\Qt d\Qt^\dagger\, e^{-\f{1}{2}\Tr
(\Qt\Qt^\dagger)^2} e^{ix\Tr(\Qt+\Qt^\dagger)}
\label{7}
\eqnx
reducing to
\eq
Z [x]= \int_0^\infty\!\!r dr\, e^{-\f{1}{2}r^4} J_0(2x r)
\label{8}
\eqx
for one flavor.
Expanding this integral in powers of $x$,
\be
  Z[m]=Z[0] \left[ 1 +\sum_{k=1}^\infty\!
	\f{2^k (m/\Delta_*)^{2k}}{(k!)^2\sqrt{\pi}} \Gamma\left(\f{k+1}2\right)
	\right]
\ee
gives rise to spectral 
sum rules for the moments of the reciprocals of the eigenvalues 
of ${\cal M}$ by matching the mass power in the spectral representation 
of the partition function,
\be
  Z[m]/Z[0]=\left\langle\prod_{\lambda_k>0}
	\left(1+\f{m^2}{\lambda_k^2}\right)\right\rangle_0
\ee
where the averaging is done over the Gaussian randomness 
with the additional measure $\prod_{\lambda_k>0} \lambda_k^2$. 
Matching the terms of order $m^2$ yields
\be
  \left\langle\sum_{\lambda_k>0} \f 1{\lambda_k^2}\right\rangle_0
	= \f{2}{\sqrt{\pi}\Delta_*^2} \,,
\label{sum1}
\ee
and matching the terms of order $m^4$ gives
\be
  \left\langle\left(\sum_{\lambda_k>0} \f {1}{\lambda_k^2}\right)^2
\right\rangle_0
	- \left\langle\sum_{\lambda_k>0} \f 1{\lambda_k^4}\right\rangle_0
	= \f 1{\Delta_*^4} \,.
\label{sum2}
\ee
The relations (\ref{sum1}-\ref{sum2}) are examples of microscopic sum rules
at the critical point $t=1$. 

One should note here that the preceding calculation has been
performed for the gaussian matrix model. It turns out that if we were
to add terms of the form 
\eq
e^{-Ng_0\Tr (PP^\dagger)^2 -Ng_1(\Tr PP^\dagger)^2}
\label{9}
\eqx
to the measure in (\ref{5}), then the spectral sum rules are 
in general unchanged. Indeed, for $N_f=1$ the addition amounts
to a global shift $x\to x/(1\!+\!2g_0\!+\!2 g_1)$. 
Higher powers of $PP^\dagger$ are subleading after rescaling 
and do not affect the sum rules in large $N$. 

\vskip 0.3cm

{\bf 4.} 
The diagonal moments of the reciprocals of the rescaled eigenvalues
are generated by the microscopic density (\ref{1}), in the limit
$N\rightarrow \infty$ and $\lambda\rightarrow 0$ but $s=\lambda/\Delta_*$
fixed. The microscopic density (\ref{1}) for $N_f$ flavours in a fixed 
topological sector $n$ : $\nu_{*,N_f,n}$, can be evaluated using 
supersymmetric methods~\cite{GENERAL,SUPER}. For example,

\eq
\hspace*{-2mm}
\nu_{*,0,0} (s)=-\f{2^{\f 14}8}{\pi^2} \big( k_2(s)j_0(s)\!+\!
	k_0(s)j_2(s)\!-\!k_1(s)j_1(s) \big) 
\label{eq:tc}
\eqx
where $k_n(s)$ and $j_n(s)$ are given by
\eqn
j_n(s)&=&\int_0^\infty\!dz\ z^{2n+1} J_0(2 z \f{s}{2^{3/4}}) e^{-z^4/2}\\ 
k_n(s)&=&\int_C\!dz\ z^{2n+2} K_1(2 z \f{s}{2^{3/4}}) e^{z^4/2}
\label{defk}
\eqnx
and where the integration contour $C$ is the sum of two lines:
$C=[-(1+i)\infty, 0] \cup [0,(1-i)\infty]$. 
In this spirit, the first sum rule (\ref{sum1}) reads
\be
\int_0^{\infty} \frac 1{s^2}\, \nu_{*,1,0} (s)\, ds = \frac{2}{\sqrt{\pi}}
	\,.
\ee
The second sum rule (\ref{sum2}) involves the 2-level microscopic correlator
for $N_f=1$ and $n=0$ which can be obtained using a similar reasoning.

\begin{figure}[tbp]
\centerline{\epsfysize=40mm \epsfbox{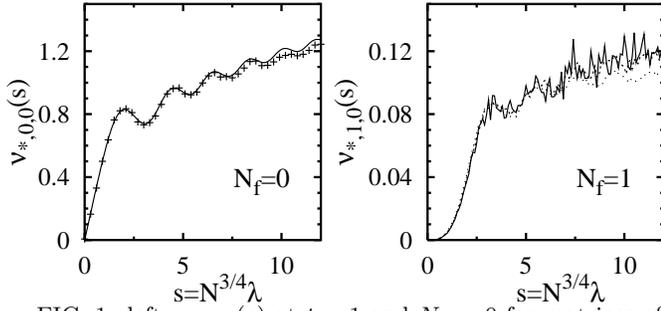}}
\caption{left: $\nu_{*,0,0} (s)$  at $t=1$ and $N_f=0$
for matrices of size N=100 (dots) and the theoretical
prediction~(\protect\ref{eq:tc}) (solid line). right: 
$\nu_{*,1,0}(s)$ at $t=1$ and $N_f=1$ for matrices of 
size $N=20$ (dotted), $N=50$ (dashed) and $N=100$ (solid).
}
\label{fig-critth}
\end{figure}

In Fig.~\ref{fig-critth} we compare the quenched $N_f=1$ (left) and unquenched
$N_f=1$ (right) results to the numerically generated microscopic spectral
density at $t=1$ using $N=100$ size matrices. 
The agreement suggests that the present method of finding the scaling
properties of microscopic spectral distributions can be used to accurately
determine the value of the critical exponent $\delta$ in lattice
simulations.

\vskip 0.3cm

{\bf 5.}
In QCD the finite temperature transition is still debated. For
very light quark masses, it is suspected to be second order, although
a cross over is not ruled out. The character of the transition depends 
crucially on the number of flavors $N_f$ and the fate of the $U_A(1)$ 
quantum breaking. For two light flavors, we may assume with Pisarski and 
Wilczek~\cite{PISARSKI} that the transition is from an $SU(2)$ spontaneously 
broken phase to $Z_2\times SU(2)\times SU(2)\sim O(4)$~\cite{NOTE}.
Choosing the order parameter $\Phi=\sigma+i\vec\tau\cdot\vec\pi$ 
(vacuum analogous to a ferromagnet), implies for the 
Ginzburg-Landau potential 

\be
{\cal V} (\Phi ) = &&+m \,\, {\rm Tr} 
(\Phi^{\dagger}+\Phi ) \nonumber\\
&& + g_0 (T) {\rm Tr} (\Phi^{\dagger} \Phi )
+ g_1 (T) ({\rm Tr} (\Phi^{\dagger} \Phi ))^2 + ...
\label{pis}
\ee
at zero vacuum angle $\theta$. The dots refer to marginal or irrelevant terms. 
For a second order transition, $g_0 (T)\sim T-T_c$, which 
is negative below $T_c$ and positive above. Note, that in this section
we analyze this hamiltonian just from the mean-field point of view by
truncating it to the space of constant modes.

For $T=T_c$ the potential (\ref{pis}) is reminiscent of the one
for the chiral random matrix model discussed above~\cite{USNJL}. If we note
that the measure on the manifold with restored symmetry is
$e^{-\beta V_3{\cal V}}\equiv e^{-V{\cal V}}$, 
we conclude that (\ref{pis})
is enough to accommodate for the level spacing
$\Delta_*=1/V^{\delta/(\delta+1)}$ with $\delta=3$ (mean-field). 
Indeed after the rescaling $x=im/\Delta_*$ we recover
(\ref{7}) with $g_1=1/2$ and the proper identification of the manifold.

For $T>T_c$ we can use (\ref{pis}) to define new 
sum rules for the quark eigenvalues in the sector with zero
winding number~\cite{SMILGA}. In particular ($N_f=2$)
\be  
\frac 1{V^2}
\left\langle\!\sum_{\lambda_k>0}\!\frac 1{\lambda_k^2}\!\right\rangle_0 
	\!\!\!\!=\!\frac{1}{2N_f}\!\!\int\limits_{0}^{2\pi}\!\!
\frac {d\theta}{2\pi} \left\langle\!\!\left({\rm Tr} 
(e^{\frac {i\theta}{N_f}} \Phi^{\dagger}\!+\!\Phi e^{-\frac {i\theta}{N_f}}\!)
\right)^2\right\rangle.
\label{high3}
\ee
The rhs measures the variance in the scalar direction on an
invariant O(4) manifold with $e^{-V{\cal V}}$ as a measure. As $T\rightarrow 
T_c$ from above, the scalar susceptibility averaged over `$\theta$-states'
(rhs of (\ref{high3})) diverges since $\sigma$ and $\pi$ become degenerate. 
These modes are the analogue of the ones originally discussed by Hatsuda and 
Kunihiro~\cite{HATSUDA} using an effective model of QCD. For the present 
case, the calculation is done readily by rescaling $\Phi\rightarrow 
\sqrt{V}\Phi$ and noting that the quartic contribution in (\ref{pis})
becomes subleading in large $V$. Hence,

\be  
\frac 1{V}
\left\langle\sum_{\lambda_k>0}\frac 1{\lambda_k^2}\right\rangle_0 = 
\frac{1}{2g_0 (T)} \,.
\label{high4}
\ee
Near $T_c$ from above it is seen to diverge as $1/(T-T_c)$ with the
critical exponent $\gamma=1$ (mean-field). We note
that above the critical temperature, the level spacing is $V^{-1/2}$,
which is intermediate between $V^{-1}$ in the vacuum and $V^{-1/4}$
in free space. Other sum rules are of course possible. 

The possibility of microscopic sum rules above the critical temperature
reflects on the assumption that the symmetry restoration in two-flavor
QCD is driven by universality. The level correlations in the Dirac spectrum 
attests to the correlations still present on the $O(4)$ manifold despite
the gap developing in the eigenvalue density. The gap is due to the fact 
that near zero virtuality the accumulation of eigenvalues is not commensurate
with the volume V. At high temperature, the quark eigenvalues are typically 
of order $T$, and both $\pi$ and $\sigma$ correlations are dissolved in 
the `plasma' upsetting the present universality arguments. In QCD we expect
this to take place at a temperature $T\sim 3T_c$~\cite{ZAHED}.

\vskip 0.3cm
{\bf 6.}
The idea of microscopic sum rules above the critical temperature may be
readily checked using an (unquenched) chiral random matrix model
with one flavor. 
Again we may use the bosonized form of the partition function
(\ref{e.last}), but now instead introduce the variable 
$y=\sqrt{N} im\sim \sqrt{N}\lm$,
and rescale $Q$ by $Q \lra \Qt=Q \sqrt{N}$. This leads to the
following expression for the partition function
\eqn
Z(y)&=& t^{2N}\int_0^\infty r dr
e^{-\left(1-\f{1}{t^2}\right) r^2} J_0(2y r ) e^{y^2}=\nonumber\\
&=& \f{\pi t^{2N}}{N}\cdot \f{t^2}{t^2-1}
e^{-\f{1}{t^2-1} y^2}
\eqnx
Expanding the partition function in powers of $y$ leads again to modified
sum rules, the simplest example being
\eq
\left\langle\sum_{\lambda_k>0}\frac 1{N\lambda_k^2}\right\rangle_0 =
\f{1}{t^2-1} 
\label{uqsumr}
\eqx
which is seen to diverge at $t=1$ as expected.

\begin{figure}[tbp]
\centerline{\epsfysize=42mm \epsfbox{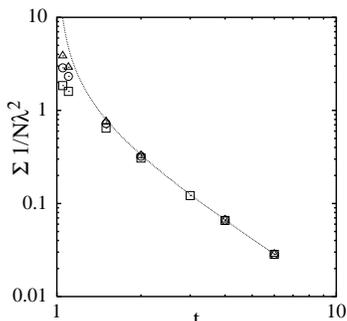}}
\caption{The result~(\protect\ref{uqsumr}) (solid line) checked against
numerical simulations using  $N=20$ (open squares), $N=50$ (open circles) and
$N=100$ (open triangles) random matrices.}
\label{fig-sumr1}
\end{figure}

The comparison to numerical simulation is shown in Fig.~\ref{fig-sumr1}
using a random set of matrices ${\cal M}$ distributed with a Gaussian measure.
For high temperatures or large matrix sizes the agreement is good.
At the critical temperature, the finite size effects are important. Amusingly,
we note the drop by two orders of magnitude at 
$t\sim 3 = 3t_c$.

\begin{figure}[tbp]
\centerline{\epsfysize=31mm \epsfbox{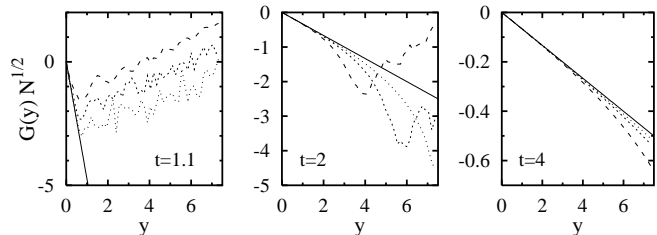}}
\caption{Scaled and quenched resolvent~(\protect\ref{qsres}) (solid line)
in comparison to a numerical simulation with $N=20$ (long dashes), 
$N=50$ (short dashes) and $N=100$ (dotted line) random matrices.}
\label{fig-qsres}
\end{figure}

The Green's function both for {\em quenched} (the mass playing the role
of an external parameter) and one flavor chiral random matrix model
in the rescaled variables may be readily obtained,
\eq
G(y)=\f{1}{\sqrt{N}} \sum_i \f{1}{y-\sqrt{N}\lm_i} = 
-\f{1}{\sqrt{N}} \f {y}{t^2-1} \,.
\label{qsres}
\eqx
Since for $T>T_c$ the eigenvalue spectrum develops a gap, there are no
eigenvalues for small values of $y$, hence the resolvent is purely
real. For finite sizes however, $y$ may get out of the gap and our
result breaks down as well as the scaling arguments. This is seen in 
Fig.~\ref{fig-qsres}, where for temperatures slightly above the critical one
we are entering the nonzero eigenvalue density part of the spectra. This 
effect is shifted for higher values with increasing temperature and matrix 
size. 

\vskip 0.3cm
{\bf 7.}
Using arguments based on universality we have suggested that the
QCD Dirac spectrum may exhibit universal spectral correlations at $T=T_c$ 
that reflect on the nature of the chirally restored phase. To illustrate
our points, we have used a chiral random matrix model.
Although the matrix model is based on a Gaussian weight, we provided 
physical arguments for why the results are insensitive to the choice of the 
weight at the critical point. In QCD this can be readily checked using our
recent arguments~\cite{QCD} at finite temperature, since the closest 
singularity to zero in the virtuality plane is persistently `pionic' 
for $T\leq T_c$~\cite{KOCIC}. 

The existence of a microscopic spectral density at $T=T_c$ for
QCD opens up the interesting possibility of measuring both the
critical temperature $T_c$ and the critical exponent $\delta$ 
by simply monitoring the pertinently rescaled distribution 
$\nu_*(s)$ of eigenvalues in lattice QCD simulation. We have also
suggested that these correlations may persist even above $T_c$.

\vskip .5cm
{\bf Acknowledgements}
\vskip .3cm

After completing this paper we noticed the paper by Brezin and Hikami
cond-mat/9804023 were similar issues are discussed using different arguments.
This work was supported in part by the US DOE grant DE-FG-88ER40388, by the
Polish Government grant Project (KBN) grants 2P03B04412 and 2PB03B00814 and 
by the Hungarian grants FKFP-0126/1997 and OTKA-T022931. RAJ is
supported by the Foundation for Polish Science (FNP).


\begin{thebibliography}{99}
\bibitem{NOTE}
The $Z_2$ factor is from the persistent $U_A(1)$ breaking effects.
\end{thebibliography}
\end{document}